\documentclass[prd,preprint,
superscriptaddress,showpacs,nofootinbib,%
tightenlines
]{revtex4}
\usepackage{epsfig}
\usepackage{color}
\usepackage{amssymb}

\newcommand{\ben}{\begin{displaymath}}
\newcommand{\een}{\end{displaymath}}
\newcommand{\be}{\begin{equation}}
\newcommand{\ee}{\end{equation}}
\newcommand{\bea}{\begin{eqnarray}}
\newcommand{\eea}{\end{eqnarray}}
\begin{document}

\title{Derivation of spontaneously broken gauge symmetry from \\ the consistency of effective field theory I:\\
Massive vector bosons coupled to a scalar field}
\author{D.~Djukanovic}
 \affiliation{Helmholtz Institute Mainz, University of Mainz, D-55099 Mainz, Germany}
\author{J.~Gegelia}
 \affiliation{Institute for Advanced Simulation, Institut f\"ur Kernphysik
   and J\"ulich Center for Hadron Physics, Forschungszentrum J\"ulich, D-52425 J\"ulich,
Germany}
\affiliation{Tbilisi State  University,  0186 Tbilisi,
 Georgia}
 \author{Ulf-G.~Mei\ss ner}
 \affiliation{Helmholtz Institut f\"ur Strahlen- und Kernphysik and Bethe
   Center for Theoretical Physics, Universit\"at Bonn, D-53115 Bonn, Germany}
 \affiliation{Institute for Advanced Simulation, Institut f\"ur Kernphysik
   and J\"ulich Center for Hadron Physics, Forschungszentrum J\"ulich, D-52425 J\"ulich,
Germany}
\date{17 March, 2018}
\begin{abstract}
We revisit the problem of deriving local gauge invariance with spontaneous symmetry breaking 
in the context of an effective field theory.
Previous derivations were based on the condition of tree-order unitarity. However, the modern point 
of view considers the Standard Model as the leading order approximation to an effective field theory. 
As tree-order unitarity is in any case violated by higher-order terms in an effective field theory, 
it is instructive to investigate a formalism which can be also applied to analyze higher-order interactions. 
In the current work we consider an effective field theory of massive vector bosons interacting with a 
massive scalar field.  We impose the conditions of generating the right number of constraints 
for systems with spin-one particles and perturbative renormalizability as well as the separation of
scales at one-loop order.
We find that the above conditions impose severe restrictions on the coupling constants of the interaction terms.
Except for the strengths of the  self-interactions of the scalar field, 
that can not be determined at this order 
{ from the analysis of three- and four-point functions}, we recover 
the gauge-invariant Lagrangian with spontaneous symmetry breaking taken in the unitary gauge as the 
leading order approximation to an effective field theory. We also outline the additional work that
is required to finish this program.

\end{abstract}




\maketitle

\section{Introduction}

The standard model (SM) is widely accepted as the established consistent theory of
the strong, electromagnetic and weak interactions \cite{Weinberg:mt}.  Invariance under Lorentz and 
local gauge SU(3)$_C\times$SU(2)$_L\times$U(1) transformations is taken as the underlying symmetry of the SM. 
Despite the tremendous success of the SM its structure  leaves some unanswered questions.
In particular, the electromagnetic and gravitational forces are long-ranged and therefore if they are indeed 
mediated by massless photons and gravitons, 
then the corresponding local Lorentz-invariant quantum field theories must be gauge
theories \cite{Weinberg:mt}. On the other hand, as the weak interaction is mediated by massive particles,
one might wonder
why it should  be described by a gauge theory with the spontaneous symmetry breaking? 
A gauge-invariant theory with the spontaneous symmetry breaking has been derived by demanding
tree-order unitarity of the S-matrix in 
Refs.~\cite{LlewellynSmith:1973ey,Cornwall:1973tb,Cornwall:1974km,Joglekar:1973hh}. 
This result could be considered as a (more or less) satisfactory answer to the above raised question, 
however, the modern point of view considers the SM as an 
effective field theory (EFT)  \cite{Weinberg:mt} which inevitably violates the tree-order 
unitarity condition at sufficiently high energies.  This motivates us to revisit the problem.

In the current work we address the issue of deriving the most general theory of massive vector bosons by 
demanding self-consistency in the sense of an EFT. The Lagrangian of an EFT consists of an
infinite number of terms, however, the contributions of non-renormalizable interactions in physical 
quantities are suppressed for energies much lower than some large scale.
Renormalizability in the sense of a fundamental theory is replaced by the
renormalizability in the sense of an EFT, i.e. that all divergences can be absorbed by 
renormalizing an infinite number of parameters of the effective Lagrangian.
Notice that the condition of  perturbative renormalizability in the sense of EFT  
is not equivalent to the condition of tree-order unitarity. 
While the tree-order unitarity implies renormalizability in the traditional sense, 
perturbative renormalizability in the sense of EFT is a much weaker condition and it does not 
imply  tree-order unitarity.
On the other hand for an EFT to be ``effective'' it is crucial that the scales are separated, i.e. 
the contributions of higher order operators in physical quantities are suppressed by powers of some large scale. 
This condition is much more restrictive than just renormalizability in the sense of EFT.  
Renormalizability alone can be achieved without introducing scalars, i.e. considering a theory 
of massive vector bosons and fermions \cite{Gegelia:2012yq}. However, in such a theory divergences 
generated from the leading order Lagrangian are removed by renormalizing the parameters of higher order 
interactions. This leaves the scales of the renormalized couplings of the higher order terms much too low 
to explain the tremendous success of the SM. 
Therefore, in what follows we analyze the constraint structure and the conditions of perturbative 
renormalizability and scale separation for  the most general Lorentz-invariant 
effective Lagrangian of massive vector bosons interacting with a scalar field. 
The performed analysis is similar to that of Ref.~\cite{Djukanovic:2010tb} but here we do not assume 
 parity conservation.

The most general Lorentz-invariant effective Lagrangian contains an infinite number of interaction
terms. It is assumed that all coupling constants of
``non-renormalizable'' interactions, i.e. terms with couplings of
negative mass-dimensions, are suppressed by powers of some large
scale. Massive vector bosons are spin-one particles and therefore they are described by Lagrangians with
constraints. To have a system with the right number of degrees
of freedom, the coupling constants of the Lagrangian have to satisfy some non-trivial relations.
Additional consistency conditions are imposed on the couplings by demanding  perturbative 
renormalizability in the sense of EFT and the separation of scales.
Restrictions on the couplings appear  because while all loop diagrams can be made finite in any 
quantum field theory if we include 
an infinite number of counter terms in the Lagrangian, it is by no means guaranteed that these counter 
terms are consistent with constraints of the theory of spin-one particles 
 and that the scale separation is not violated.

The paper is organized as follows: In section~\ref{VMS} we specify the effective Lagrangian and 
carry out the analysis of the constraints. The conditions of perturbative renormalizability are obtained 
in section~\ref{PRen}. 
We summarize and discuss the obtained results in section~\ref{Concl}.

\section{Constraint analysis of an EFT  Lagrangian of massive vector bosons and a scalar}

\label{VMS}

   We start with the most general Lorentz-invariant effective Lagrangian of a scalar and three massive 
vector boson fields respecting  electromagnetic charge conservation (even though we do not consider
the explicit coupling to the U(1) gauge field in the following). 
Two charged vector particles are represented by 
vector fields $V^\pm_\mu=(V^1_\mu\mp iV^2_\mu)/\sqrt{2}$, 
the third component, $V^3_\mu$, and the scalar field $\Phi$ are charge-neutral. The effective 
Lagrangian contains an infinite number of interaction terms and hence depends on an 
infinite number of parameters. We assume that coupling constants with negative mass dimensions are independent 
from those of positive and zero mass dimensions. Below we analyse the Lagrangian containing only 
interaction terms with coupling constants of non-negative dimensions (as explained in detail below).
Thus, the effective Lagrangian  under consideration can be written as follows:  
\begin{eqnarray}
{\cal L} & = & -{1\over 4} \ V^a_{\mu\nu} V^{a \mu\nu}
+\frac{M_a^2}{2} V_\mu^a V^{a \mu} - g_V^{abc} V^a_\mu V^b_\nu \partial^\mu V^{c \nu} \nonumber\\
&-&  g_A^{abc}\, \epsilon^{\mu\nu\alpha\beta}  V^a_\mu V^b_\nu \partial_\alpha V^c_\beta - h^{a b c d} V_\mu^a V_\nu^b V^{c \mu}V^{d\nu}
\nonumber\\
& + & \frac{1}{2}\, \partial_\mu\Phi \,\partial^\mu\Phi -\frac{m^2}{2}\, \Phi^2  -a\,\Phi -\frac{b}{3!}\,\Phi^3- \frac{\lambda}{4!}\,\Phi^4
\nonumber\\
& - &  g_{vss} \,\partial_\mu V^{3\mu} \Phi^2 - g^{ab}_{vvs} \,V^{a\mu}\,V^b_\mu \Phi - g^{ab}_{vvss} \,V^{a\mu}\,V^b_\mu \Phi^2\,,
\label{Lagr}
\end{eqnarray}
where $V^a_{\mu\nu}=\partial_\mu V^a_\nu -
\partial_\nu V^a_\mu $, $M_a$ are vector boson masses ($M_1=M_2=M$), $m$ is the mass of the scalar
 and the summation over all repeated indices runs from 1 to 3. Note that since the vector bosons
are not related to some gauge symmetry, the three- and four-boson couplings have independent 
coupling constants.
We did not include the mixed term $\partial_\mu V^{3\mu} \Phi $ as it can be eliminated by a suitable 
field redefinition. Due to the  electromagnetic charge conservation not all coupling constants of the above Lagrangian 
are independent from each other. 
The interaction terms of the scalar field with two vector fields can be written in terms of four real parameters
\begin{equation}
g_{1,s} = g^{11}_{vvs} = g^{22}_{vvs} , \ \  g_{2,s} = g^{33}_{vvs}, \ \  g_{1,ss} = g^{11}_{vvss} = g^{22}_{vvss}, \ \  g_{2,ss} = g^{33}_{vvss},
\label{pVVCs}
\end{equation}
and all other $g^{ab}_{vvs}$  and  $g^{ab}_{vvss}$ couplings do not contribute in the effective Lagrangian. 
The coefficient of the linear term $a$ vanishes at tree order and further corrections can be 
fixed by demanding that the vacuum expectation value (vev) of the scalar field vanishes.

The three-boson interaction term of the Lagrangian depends on ten real parameters,
\begin{eqnarray}
g_V^{333} & = & g_1, \ \ \ g_V^{113}= g_2, \ \ \ g_V^{123}= -g_3, \ \
\ g_V^{213}= g_3,\nonumber\\
g_V^{223} & = & g_2, \ \ \ g_V^{311}= g_4, \ \ \ g_V^{321}= -g_5, \ \ \
g_V^{312}= g_5, \nonumber\\
g_V^{322} & = & g_4, \ \ \ g_V^{131}= g_6, \ \ \ g_V^{231}= -g_7, \ \ \
g_V^{132}= g_7,\ \ \ g_V^{232} =  g_6\,,\nonumber\\
g_A^{213} & = & -g_A^{123}= g_{A1},\ \ \
g_A^{311}  =  g_A^{322}= -g_A^{131}= -g_A^{232}= g_{A2}, \nonumber\\
g_A^{312} & = & -g_A^{321}= -g_A^{132}= g_A^{231}= g_{A3} \,.
\label{3couplings}
\end{eqnarray}
 Electromagnetic charge conservation relates  the coupling constants of the four-boson interaction $h^{a b c d}$ 
to each other as follows 
\begin{eqnarray} h^{1111} & = & h^{2222}=\frac{d_1+d_2}{4}\,, \nonumber\\
h^{1112} & = & -h^{1121}-h^{1211}-h^{2111},\nonumber\\
h^{1122} & = & d_2-h^{2112}-h^{1221}-h^{2211},\nonumber\\
h^{1212} & = & \frac{1}{2}\left(d_1-d_2-2\,h^{2121}\right),\nonumber\\
h^{1323} & = & -h^{2313}-h^{3132}-h^{3231},\nonumber\\
h^{2122} & = & -h^{1222}-h^{2212}-h^{2221},\nonumber\\
h^{2323} & = & \frac{1}{2}\left(d_4-2\,h^{3232}\right),\nonumber\\
h^{3113} & = & \frac{1}{2}\left[d_3-2\,\left(h^{1133}+h^{1331}+h^{3311}\right)\right],\nonumber\\
h^{3223} & = & \frac{1}{2}\left[d_3-2\,\left(h^{2233}+h^{2332}+h^{3322}\right)\right],\nonumber\\
h^{3123} & = & -h^{1233}-h^{1332}-h^{2133}-h^{2331} -  h^{3213}-h^{3312}-h^{3321},\nonumber\\
h^{3131} & = & \frac{1}{2}\left(d_4-2\,h^{1313}\right),\nonumber\\
h^{3333} & = & d_5\,, \label{4couplingsNEW}
\end{eqnarray}
and the effective Lagrangian of Eq.~(\ref{Lagr}) depends only on $d_1,\cdots,d_5$.

\medskip


Details of the canonical formalism followed below can be found in Ref.~\cite{gitman}. 
Our analysis is closely related to that of Ref.~\cite{Gegelia:2012yq}, 
which considered an EFT without the scalar field, and it is similar to the one of Ref.~\cite{Djukanovic:2010tb}, 
with the difference that in Ref.~\cite{Djukanovic:2010tb}, parity conservation has been taken as an input.

The canonical momenta corresponding to $\Phi$, $V^a_0$ and $V^a_i$ are defined as
\begin{eqnarray}
p &=& {\partial {\cal L}\over \partial\dot \Phi }= \dot \Phi\, \label{p},
\\
\pi^a_0 &=& {\partial {\cal L}\over \partial\dot V^a_0}=-g_V^{bca}
V^b_0 V^c_0-g_{vss} \delta_{a3}\,\Phi^2 
\label{pi0},
\\
\pi^a_i &=& {\partial {\cal L}\over \partial\dot V^a_i}=V^a_{0i}
+g_V^{bca} V^b_0 V^c_i+g_A^{bca}\epsilon^{ijk0} V^b_j V^c_k\,. \label{pii}
\end{eqnarray}
Eq.~(\ref{pi0}) leads to the primary constraints
\begin{equation}
  \phi_1^a=\pi^a_0+g_V^{bca} V^b_0 V^c_0 +g_{vss}\, \delta_{a3}\,\Phi^2 \,.
  \label{phi1}
\end{equation}
On the other hand, from Eqs.~(\ref{p}) and (\ref{pii}) we solve
\begin{eqnarray}
\dot V_i^a & = & \pi_i^a+\partial_iV_0^a-g_V^{bca}V_0^b V_i^c
- g_A^{bca}\epsilon^{ijk0} V^b_j V^c_k\,,\nonumber\\ 
\dot\Phi & = & p\,.
\label{aidot}
\end{eqnarray}
For the total Hamiltonian \cite{gitman} we have:
\begin{equation}
H_1=\int d^3{\bf x}\, \left( \phi^a_1 z^a +{\cal H}\right) 
 \label{hamden}
\end{equation} 
with
\begin{eqnarray}
{\cal H} & = & {\pi^{a}_i\pi^{a}_i\over 2} +\pi_i^a
\partial_i V^a_0 + \frac{1}{4} V^a_{ij}
V^a_{ij} - {M_a^2\over 2}V^a_\mu V^{a \mu}
-g_V^{abc}V^a_0 V^b_i \pi^c_i - g_A^{abc}\epsilon^{ijk0}V^a_j V^b_k \pi^c_i\nonumber\\
&-& g_V^{abc} V_0^a V_i^b
\partial_iV^c_0-g_V^{abc} V_i^a V_0^b
\partial_iV^c_0
+ g_V^{abc} V_i^a V_j^b \partial_i V^c_j+\frac{1}{2} g_V^{abc}
g_V^{a'b'c} V_0^a V_i^b
V^{a'}_0 V_i^{b'} \nonumber\\
&+& \frac{1}{2} g_A^{bca}
g_A^{b'c'a} \epsilon^{ijk0}\epsilon^{i j' k' 0}\,V_j^b V_{j'}^{b'} V^{c}_k V_{k'}^{c'}
+ g_A^{abc}
g_V^{b'c'c} \epsilon^{ijk0}\,V_j^a V_{k}^b V^{b'}_0 V_{i}^{c'}\nonumber\\
&+&g_A^{abc} \epsilon^{ijk0} V^a_0 V^b_j \partial_i V_k ^c-g_A^{abc} \epsilon^{ijk0} V^a_j V^b_0
\partial_i V_k ^c
+ h^{abcd} V_\mu^a V_\nu^b V^{c \mu} V^{d \nu}\nonumber\\
& + & \frac{p^2}{2}+\frac{\partial_i\Phi^2}{2} + \frac{m^2}{2}\, \Phi^2  + a\,\Phi + \frac{b}{3!}\,\Phi^3 + \frac{\lambda}{4!}\,\Phi^4
\nonumber\\
& + &  g_{vss} \,\partial_i V^{3 i} \Phi^2 + g^{ab}_{vvs} \,V^{a\mu}\,V^b_\mu \Phi + g^{ab}_{vvss} \,V^{a\mu}\,V^b_\mu \Phi^2
\,,
\label{hamiltonian}
\end{eqnarray}
and the $z^a$ are arbitrary functions which must be determined.

The primary constraints $\phi^a_1$ have to be conserved in time, i.e. their Poisson 
brackets with the Hamiltonian must vanish for each $a=1,2,3$. Calculating the 
Poisson brackets we obtain
\begin{eqnarray}
\left\{ \phi^a_1,H_1\right\}&=&
\left(g_V^{bca}+g_V^{cba}-g_V^{acb}-g_V^{cab}\right) V_0^c z^b
+ \partial_i\pi_i^a+g_V^{abc}V_i^b\pi_i^c+
\left( g_V^{abc}+g_V^{bac}\right)\,V^b_i\partial_iV^c_0\nonumber\\
& - & g_V^{bca}\partial_i\left( V^b_0V^c_i\right) -
g_V^{cba}\partial_i\left( V^b_0V^c_i\right)+ M_{a}^2 V^a_0 
- g_V^{abc} g_V^{a'b'c} V_i^b V^{a'}_0 V_i^{b'} \nonumber\\ &-&
g_A^{a'bc} g_V^{ac'c} \epsilon^{ijk0}\,V_j^{a'} V_{k}^b V_{i}^{c'}
-g_A^{abc} \epsilon^{ijk0} V^b_j \partial_i V_k ^c+g_A^{b a c} \epsilon^{ijk0} V^{b}_j
\partial_i V_k ^c
\nonumber\\
&-& \left( h^{abcd}+ h^{badc}+ h^{cbad}+ h^{dcba}\right)
V_\mu^b V^c_0 V^{d \mu}
\nonumber\\
& & + 2 \, g_{vss}\delta_{a3} \, p\,\Phi - 2 g^{ab}_{vvs} \,V^b_0 \Phi -2 g^{ab}_{vvss} \,V^b_0 \Phi^2
\equiv A^{ab}z^b+\chi^a. \label{equivphi2}
\end{eqnarray}
The $3\times 3$ matrix $A$ is given by
\begin{equation}
\label{defA} A=\left(\begin{array}{ccc} 0&-2\gamma_1 V_0^3&
\gamma_2 V_0^1-\gamma_1 V_0^2\\
2\gamma_1V_0^3 & 0 & \gamma_1 V_0^1+\gamma_2 V_0^2\\
-(\gamma_2 V_0^1-\gamma_1 V_0^2)& -(\gamma_1 V_0^1+\gamma_2 V_0^2)&0
\end{array}\right),
\end{equation}
where $\gamma_1=g_5+g_7$ and
$\gamma_2=g_4+g_6-2g_2$.
The determinant of $A$ vanishes and therefore the system of equations
\begin{equation}
A^{a b} z^b= -\chi^a \label{equations1}
\end{equation}
can be satisfied only if the right-hand side satisfies the secondary constraint
\begin{equation}
\phi_2= \chi^1\,(\gamma_1 V_0^1+\gamma_2 V_0^2) + \chi^2\, (\gamma_1
V_0^2-\gamma_2 V_0^1)-\chi^3\, 2 \gamma_1\, V_0^3 = 0\,.
\label{additionalconstraint1}
\end{equation}
If  at least one of $\gamma_1$ or $\gamma_2$ is non-zero then for non-vanishing  
$V_0^1$ and/or $V_0^2$ we obtain from Eq.~(\ref{equations1}) that
\begin{eqnarray}
z^1 & = & \frac{\chi_3+\gamma_1 z^2\,V_0^1+\gamma_2 \,z^2
V_0^2}{\gamma_1\, V_0^2-\gamma_2\, V_0^1},\nonumber\\
z^3 & = & \frac{\chi_1+2\,
\gamma_1\,z^2\,V_0^3}{\gamma_2\,V_0^1-\gamma_1\,V_0^2}
\label{zsolutions}
\end{eqnarray}
and $z^2$ can be solved from time conservation of the constraint 
$\phi_2$, $\left\{ \phi_2,H_1\right\}= 0$.
   However, in this case we obtain four constraints 
of the second class instead of six  for our system of three
massive vector fields.
   Therefore, for a self-consistent theory we must require
\begin{equation}
\gamma_1=\gamma_2=0\Rightarrow \ g_7= - g_5\,, \ \  2g_2=g_4+g_6\,. \label{gfirstconstraint}
\end{equation}
Thus we are left with secondary constraints:
\begin{eqnarray}
\left\{ \phi^a_1,H_1\right\}&=&
\partial_i\pi_i^a+g_V^{abc}V_i^b\pi_i^c+
\left( g_V^{abc}+g_V^{bac}\right)\,V^b_i\partial_iV^c_0
-g_V^{bca}\partial_i\left( V^b_0V^c_i\right) -
g_V^{cba}\partial_i\left( V^b_0V^c_i\right)+ M_{a}^2 V^a_0 \nonumber\\ &-&
 g_V^{abc} g_V^{a'b'c} V_i^b V^{a'}_0 V_i^{b'} 
- g_A^{a'bc}
g_V^{ac'c} \epsilon^{ijk0}\,V_j^{a'} V_{k}^b V_{i}^{c'}
-g_A^{abc} \epsilon^{ijk0} V^b_j \partial_i V_k ^c+g_A^{b a c} \epsilon^{ijk0} V^{b}_j
\partial_i V_k ^c 
\nonumber\\
&-&   \left( h^{abcd}+ h^{badc}+ h^{cbad}+ h^{dcba}\right)
V_\mu^b V^c_0 V^{d \mu}\nonumber\\
&+& 2 \, g_{vss}\delta_{a3} \, p\,\Phi - 2 g^{ab}_{vvs} \,V^b_0 \Phi -2 g^{ab}_{vvss} \,V^b_0 \Phi^2
\equiv \phi_2^a,\quad a=1,2,3.
\label{equivphi22}
\end{eqnarray}
If no more constraints appear then our
Lagrangian describes a system with the right number of constraints for three massive vector bosons 
interacting with a scalar particle. 
If this is the case, then the $z^a$ have to be solvable from the
condition of the constraints $\phi_2^a$ being conserved in
time. 

  From the condition of  conservation of $\phi^a_2$ in time we obtain
\begin{equation}
 \left\{ \phi^a_2,H_1\right\} = {\cal M}^{ab} z^b +Y^a =
0,\quad a=1,2,3,
\label{zaequations}
\end{equation}
where 
\begin{eqnarray}
{\cal M}^{ab} & = & M^2_a\delta^{ab}
-\left(g_V^{bca}+g_V^{cba}\right)\partial_i V_i^c - \bigl[ g_V^{ace} g_V^{bde} 
- \left( h^{acbd}+ h^{cadb}+ h^{bcad}+ h^{dbca}\right) \bigr] V_i^c V_i^d\nonumber\\
&-& \bigl(h^{abcd} +h^{badc} + h^{cbad}+h^{dcba} +h^{adcb}
+h^{dabc}+h^{cdab} +h^{bcda} + h^{acbd} +h^{cadb} \nonumber\\
& + & h^{bcad} +h^{dbca}\bigr) V_0^c V_0^d-4\, g_{vss}^2 \delta_{a3} \delta_{b3} \Phi^2-2\, g_{vvs}^{ab} \Phi-2\, g_{vvss}^{ab} \Phi^2,
\label{commutmatr}
\end{eqnarray}
and the  particular form of $Y^a$ is not important for our purposes. 
To obtain a self-consistent field theory we demand
that ${\rm det} {\cal M}$ does not vanish. 
For small fluctuations about the vacuum this is indeed the case and we proceed by 
quantizing these small fluctuations and deriving further constraints on the couplings by 
investigating the conditions of perturbative renormalizability and scale separation. 

\section{Perturbative renormalizability}
\label{PRen}

Below we analyze one-loop order diagrams using  dimensional regularization 
(see, e.g., Ref.~\cite{Collins:1984xc}). 
To that end after lengthy calculations we obtain the following generating functional 
\begin{equation}
Z[J^{a\mu},I] = \int {\cal D} V\,{\cal D} \Phi \,{\cal D}\,c\,{\cal D}\,\bar c\,{\cal D}
\lambda\,e^{i \int d^4 x \,\left( {\cal L} + {\cal L}_{\rm aux}(\bar c, c,  \lambda, \Phi, V)+J^{a\mu}V_\mu^a +I \Phi\right)},
\label{GFPEffectiveCanonicalcoordinates}
\end{equation}
where the particular form of ${\cal L}_{\rm aux}(\bar c, c, \Phi, V \lambda)$ is unimportant as 
it generates vanishing contributions to Feynman diagrams if  dimensional regularization is applied. 
This is because the $\lambda^a$, $c^a$ and $\bar c^a$ fields do not have kinetic parts. In 
the calculations of the loop diagrams below we used the programs
FeynCalc \cite{Mertig:1990an} and Form \cite{Kuipers:2012rf}
independently. The divergent parts of the one-loop integrals have been checked with the
expressions obtained in  Ref.~\cite{Denner:2005nn}.

We impose the on-mass-shell renormalization condition, i.e. require that all divergences in 
physical quantities should be removable by redefining the parameters of the effective Lagrangian.

\begin{figure}[t!]
\epsfig{file=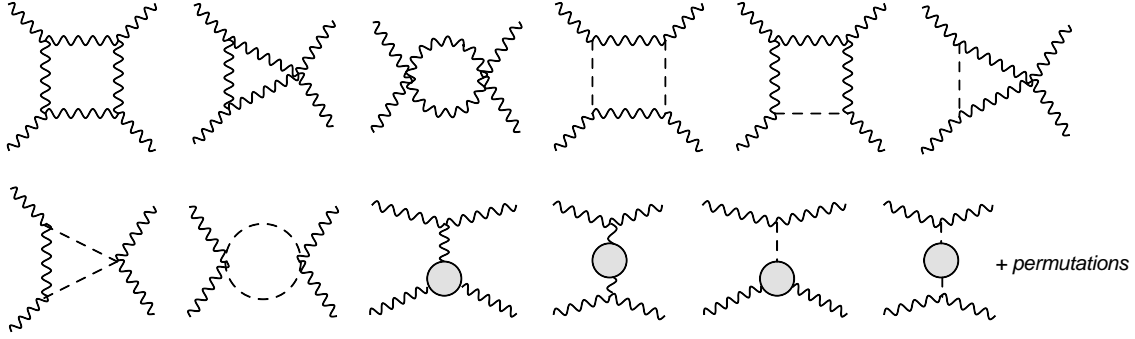, width=15truecm}
\caption[]{\label{VVVV:fig} One-loop contributions to the
four-vector vertex function. The dashed and the wiggly lines correspond to the
scalar and the vector-boson, respectively. Blobs indicate the corresponding one-loop two- and 
three-point vertex functions. In last four diagrams only the one-particle-irreducible parts are 
taken into account.}
\end{figure}

\begin{figure}[t!]
\epsfig{file=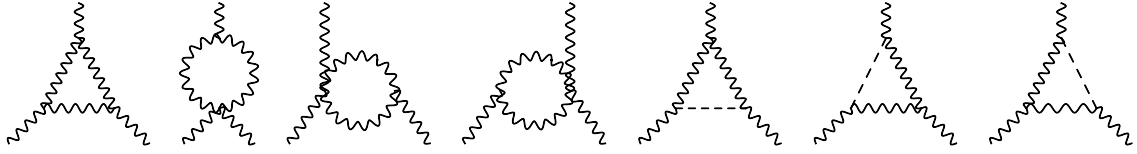, width=15truecm}
\caption[]{\label{VVV:fig} One-loop contributions to the
three-vector vertex function. The dashed and wiggly lines correspond to the
scalar and the vector-boson, respectively.}
\end{figure}

\begin{figure}[t!]
\epsfig{file=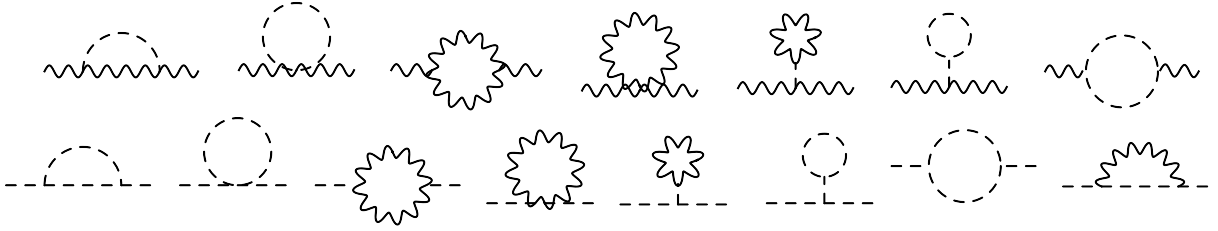, width=16truecm}
\caption[]{\label{SE:fig} One-loop contributions to the
 vector-boson and the scalar self-energies. The first and second lines represent the vector boson and scalar self-energies. The dashed and the wiggly lines correspond to the
scalar and the vector-boson, respectively. }
\end{figure}

\begin{figure}[t!]
\epsfig{file=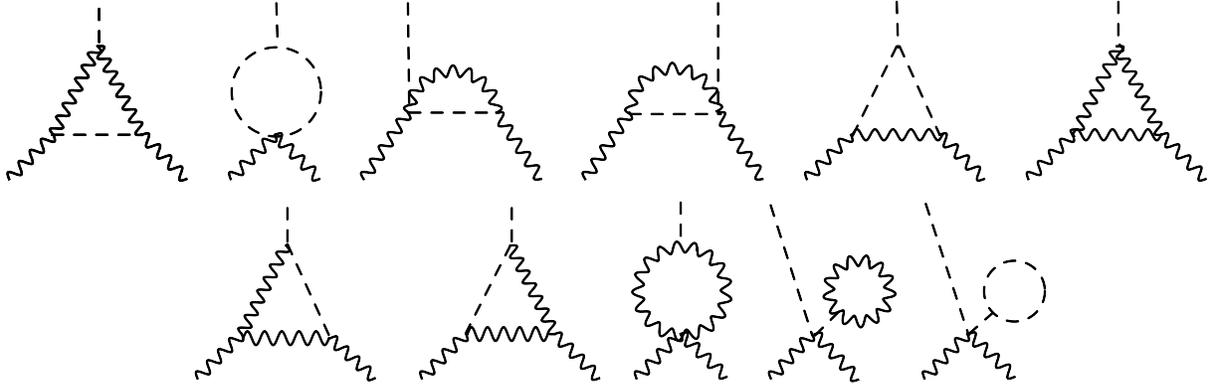, width=16truecm}
\caption[]{\label{SVV:fig} One-loop contributions to the
scalar-vector-vector vertex function. The dashed and the wiggly lines correspond to the
scalar and the vector-boson, respectively. }
\end{figure}

We start by calculating the  one-loop contribution to the scattering amplitude  $V^3V^3\to V^3V^3$,
shown in Fig.~\ref{VVVV:fig}.  The coefficient of the divergence is a polynomial of the Mandelstam 
variables ($s,t,u$) divided by powers of the vector boson masses. 
If these divergences are removed by renormalizing the coupling constants of the higher order 
operators, i.e. four-vector interaction terms with derivatives, then the scale of 
these couplings will be set by the masses of vector bosons. This would mean that the contributions of 
higher order operators in physical quantities would not be suppressed by powers of a large 
scale (but rather by the vector boson mass divided by some power of a dimensionless coupling constant).  
Therefore to have a self-consistent perturbative EFT with clear scale-separation divergences generated 
by interactions with dimensionless (or positive mass dimension-) couplings should either cancel 
each other or be removable by renormalizing this set of leading order couplings. 
Notice here that non-pole parts of  one-particle reducible diagrams in Fig.~\ref{VVVV:fig} have to 
be taken into account together with one-particle irreducible diagrams. The dressed vertices and 
self-energies of the scalar and vector particles contributing in one-particle reducible diagrams 
are given by respective diagrams in Figs.~\ref{VVV:fig},  \ref{SE:fig} and \ref{SVV:fig}.\footnote{In 
all figures only those diagrams are shown which contribute to the divergent parts of the corresponding 
scattering amplitudes.}  
We do not give very lengthy expressions of the divergent parts of loop diagrams but rather 
show only the conditions imposed on the coupling constants. 

 Consider first  the term proportional to $u^4$  in the coefficient of $g^{\mu\nu}g^{\lambda\sigma}$, where 
$\mu, \nu, \lambda$ and $\sigma$  are Lorentz indices corresponding to external vector lines, and 
demand that it vanishes. This leads to
\begin{equation}
g_1^4 M_3^8+2 g_2^4 M^8=0,\Rightarrow g_1=0, \ g_2=0\,.
\label{g1g2}
\end{equation} 
The next condition is obtained by demanding that the term proportional to $u^2$ also vanishes.
This leads to the following condition on certain couplings: 
\begin{eqnarray}
&& \left[\frac{1}{4} \left(-20
   \left(g_{\text{A1}}+g_{\text{A3}}\right)^2-20
   g_{\text{A2}}^2-5 g_3^2-10 g_5
   g_3-g_4^2-g_5^2\right)+d_3\right]^2\nonumber\\
&&   +\frac{1}{16}
   \biggl[ 
   40 \left(44 g_{\text{A2}}^2+96 g_3^2+11 g_4^2+11
   \left(g_3+g_5\right){}^2\right)
   \left(g_{\text{A1}}+g_{\text{A3}}\right)^2 +\frac{32  M^4}{M_3^4} \, d_5^2 \nonumber\\
&& +880
   \left(g_{\text{A1}}+g_{\text{A3}}\right){}^4+55 \left(4
   g_{\text{A2}}^2+g_4^2+\left(g_3+g_5\right){}^2\right)^2
   \biggr]
   =0 \,. \label{condition2}
\end{eqnarray}
Eq.~(\ref{condition2}) leads to
\begin{eqnarray}
d_5 & = & 0,\nonumber\\ 
g_4 & = & 0,\nonumber\\
g_5 &=& - g_3,\nonumber\\
g_{A2} &=& 0,\nonumber\\
g_{A3} &=& - g_{A1},\nonumber\\
d_3 &=&-g_3^2. \label{d3g4}
\end{eqnarray}
Further, demanding the vanishing of the term proportional to $s^2$, we obtain
\begin{equation}
d_4=g_3^2\,.
\label{d4}
\end{equation} 
\noindent
Taking into account Eqs.~(\ref{g1g2}), (\ref{d3g4}) and (\ref{d4}) the full expression of the divergent part of 
the amplitude $V^3V^3\to V^3V^3$ becomes proportional to
\begin{equation}
 \left[8 M^8 \left(2 M_3^2
   g_{2,\text{ss}}+g_{2,s}^2\right){}^2+M_3^4 \left(g_3^2
   M_3^4-4 M^2 g_{1,s} g_{2,s}\right){}^2\right]
   \left(g^{\lambda \sigma } g^{\mu \nu }+g^{\lambda \nu }
   g^{\mu \sigma }+g^{\lambda \mu } g^{\nu \sigma
   }\right). 
\label{xx2}
\end{equation}

For $d_5=g_1=0$, the one-particle-irreducible tree-order contribution to $V^3V^3\to V^3V^3$ amplitude 
vanishes and therefore we have to demand that the expression in Eq.~(\ref{xx2}) also vanishes. 
Doing so we  obtain:
\begin{eqnarray}
g_{2,\text{ss}} & = & -\frac{g_3^4 M_3^6}{32 M^4 g_{1,s}^2},\nonumber\\
g_{2,s} &=& \frac{g_3^2 M_3^4}{4 M^2 g_{1,s}}~. \label{gss}
\end{eqnarray}

\noindent
Next, as there is no tree order one-particle irreducible contribution in the amplitude 
$V^1V^1\to V^1V^1$, we have to demand that the divergent part
of the corresponding one-loop contribution vanishes (diagrams shown in Fig.\ref{VVVV:fig}). 
By demanding that the terms proportional to $s^2$ and $s \,t$ vanish, we obtain the 
following conditions:
\begin{eqnarray}
&& (d_1 + d_2) \left(
   d_2+ \frac{g_3^2}{2}\right)  =0 \,, \nonumber\\
&& \left(d_2+\frac{g_3^2}{2}\right)^2+\left(d_1+d_2\right){}^
   2+\frac{1}{4} \, g_3^4 \left(1-\frac{M^4}{M_3^4}\right)=0  \,.
\label{eqx}
\end{eqnarray}
Considering the amplitude of the scalar boson decaying into two vectors and requiring 
that the divergences of corresponding diagrams, shown in Fig.~\ref{SVV:fig}, do not contribute
in the renormalization of the couplings of the higher-order operators, i.e. 
that they do not violate the scale separation, we find that the following condition has 
to be satisfied:   
\begin{equation}
g_{1,s}\left( (d_1+d_2)+d_2+\frac{g_3^2}{2}\right)(m^2-10 M^2) =0\,.
\label{svvc}
\end{equation}
The coupling $g_{1,s}$ cannot be vanishing due to the condition of Eq.~(\ref{gss}) 
and therefore from Eqs.~(\ref{eqx}) and (\ref{svvc}) we obtain
\begin{equation}
d_1 = -d_2 = \frac{g_3^2}{2} \,, \ \ M_3=M\,.
\label{solx}
\end{equation}

Using all conditions imposed on couplings so far and analyzing the vertex function $V_1V_2V_3$ 
and demanding that the divergent part of the sum of loop diagrams, shown in Fig.~\ref{VVV:fig},  
has the same  Lorentz structure as the tree one, we obtain 
 \begin{equation}
   g_{1,s}= g_{2,s}\,.
\label{g1sg2s}
\end{equation}
Eqs.~(\ref{gss}) and (\ref{g1sg2s}) lead to 
 \begin{equation}
   g_{1,s}= g_{2,s}=\pm \frac{g_3 M}{2}\,.
\label{g1sg2sE}
\end{equation}
Going back to the $V_1V_1\to V_1V_1$ amplitude and taking into account Eqs.~(\ref{solx}) and (\ref{g1sg2sE}), 
the condition of the vanishing of its divergent part reduces to
\begin{equation}
\left(8 g_{1,\text{ss}}+
   g_3^2\right)^2 =0\,,
\label{cond1}
\end{equation}
from which we obtain
\begin{equation}
g_{1,\text{ss}}=-\frac{
   g_3^2}{8} \,.
\label{solg1ss}
\end{equation}

 Next, we have  calculated the divergent parts of one-loop diagrams contributing to 
the $\Phi V_3\to \Phi V_3 $ scattering amplitude. As the coupling constant of the 
$V^3_\mu V^{3\mu} \Phi^2$ interaction term is given by $g_{2,\text{ss}}=-g_3^2/8 $, i.e. 
in terms of the coupling of the three-vector and four-vector interaction terms, the
divergent pieces of the corresponding amplitudes have to be correlated. 
In a self-consistent theory the renormalized value for the coupling $g_3$ should be independent 
from the process that was used to fix it.
 After a lengthy one-loop calculation we found that this consistency condition requires 
that the coupling $g_{vss}$ has to vanish.

We checked in explicit calculations that all one-loop divergences appearing  in processes with 
three and four particles are absorbed in a redefinition of the coupling constants and the 
masses and no further conditions on the couplings are obtained.

\medskip

To summarize, all obtained relations among couplings and masses can be
written as
\begin{eqnarray}
M_1 & = & M_2=M_3=M \,,\nonumber\\
g_V^{abc} & = & -g_3 \,\epsilon^{abc} \,,\nonumber\\
g_A^{abc} & = & g_{A1} \,\epsilon^{abc} \,,\nonumber\\
h^{abcd} & = &
\frac{1}{4}\,g_V^{abe}g_V^{cde}\,,\nonumber\\
g_{vss} & =& 0 \,,\nonumber\\
g_{1,s} & = & g_{2,s}=\frac{g_3 M}{2},\nonumber\\
g_{1,ss} & = & g_{2,ss} =-\frac{g_3^2}{8}\,.\label{ccconditions1}
\end{eqnarray}
The sign of the couplings $g_{1,s}$ and $g_{2,s}$ can be changed to the opposite by redefining the scalar field.
We have chosen the positive sign displayed above.

For the couplings in Eq.~(\ref{ccconditions1}) 
the effective Lagrangian 
can be written in a compact form,  denoting $g_3=g$, 
\begin{eqnarray}
{\cal L} & = & -{1\over 4} \ G^a_{\mu\nu} G^{a
\mu\nu} +\frac{1}{2} \,V_\mu^a V^{a \mu}\left( M- \frac{g}{2}\,\Phi \right)^2
-g_{A 1}\epsilon^{abc}\, \epsilon^{\mu\nu\alpha\beta}  V^a_\mu
V^b_\nu
\partial_\alpha V^c_\beta\,, 
\nonumber\\
& + & \frac{1}{2}\, \partial_\mu\Phi \,\partial^\mu\Phi -\frac{m^2}{2}\, \Phi^2  -a\,\Phi -\frac{b}{3!}\,\Phi^3- \frac{\lambda}{4!}\,\Phi^4\,,
\label{EffLagr}
\end{eqnarray}
where
\begin{equation}
G^a_{\mu\nu}=V^a_{\mu\nu}-g\,\epsilon^{abc}\,V^b_\mu V^c_\nu\,.
\label{gdefinition}
\end{equation}
This Lagrangian coincides with the SU(2) locally gauge invariant Lagrangian of scalars and vector 
bosons with spontaneous symmetry breaking in the unitary gauge except for the self-interaction 
terms of the scalars. Note in particular the gauge-type form of the vector boson field strength. 
We checked by explicit calculations that no further constraints on couplings are generated by the 
condition of perturbative renormalizability of the three- and four-point functions of scalar and vector 
bosons. This leaves the two scalar self-interaction couplings unfixed. We expect that the investigation 
of the one-loop diagrams contributing in six-point functions and/or two-loop order analysis of three- 
and four-point functions will fix  these couplings such that the Lagrangian with spontaneous symmetry 
breaking taken in unitary gauge results as an unique self-consistent EFT of a massive scalar and massive vector bosons.

\section{Summary and discussions}
\label{Concl}

In the current work we revisited the problem of the uniqueness of a theory with spontaneously broken gauge 
symmetry  as a consistent framework for describing the electroweak interactions. 
Following the modern point of view of the Standard Model being the leading order approximation of 
an effective field theory we analyzed the most general Lorentz-invariant leading order 
effective Lagrangian of massive vector bosons interacting with a massive scalar field. Here,
 under leading order we mean interaction terms with couplings of non-negative mass dimensions.   

Massive spin-one particles are described by  theories with constraints. The interaction terms of 
the effective Lagrangian have to be consistent with the constraints so that the theory describes 
the dynamics of the right number of degrees of freedom. Using the standard canonical formalism, 
we analyzed the constraint structure of our effective Lagrangian and obtained consistency conditions 
which must be satisfied by the various coupling constants. Further conditions are obtained by requiring 
perturbative renormalizability.  In particular, using dimensional regularization we calculated 
the divergent parts of one-loop Feynman diagrams contributing to various physical quantities and 
analyzed the conditions of renormalizability.

  By applying dimensional regularization we can keep track of only logarithmic divergences. 
However, this is sufficient as we are looking for necessary conditions of perturbative renormalizability. 
We imposed the condition that  all logarithmic divergences generated by the interaction terms of 
the leading order effective Lagrangian should be removable from physical quantities in such a way that 
the perturbative contributions of higher-order operators remain suppressed by large scales. These 
conditions impose severe restrictions on the coupling constants such that we end up with the Lagrangian 
of spontaneously broken gauge symmetry in unitary gauge except that the coupling constants of 
the self-interactions of the scalar 
field remain unfixed. These  are not pinned down by the analysis of the UV divergences of all 
three- and four-point functions at one-loop order. We expect that the condition of perturbative 
renormalizability for three- and four-point functions at two-loop order or/and one-loop order 
amplitudes with more external legs will fix these two free couplings such that the Lagrangian 
with spontaneously broken SU(2) gauge symmetry taken in unitary gauge appears as an unique 
leading-order Lagrangian of a self-consistent EFT of a massive scalar interacting with massive vector 
bosons.  As it is well known, the S-matrix  generated by such a Lagrangian is ultraviolet finite 
being identical to the one of the renormalizable gauge  \cite{'tHooft:1971fh}. 
Extending our analysis to two-loop  calculations together with the inclusion of  
the electromagnetic interaction 
(analogously to Ref.~\cite{Djukanovic:2005ag})  and fermions  is 
relegated to forthcoming publications. 

\acknowledgments

 We thank J\"urg Gasser for useful remarks on the manuscript. 
This work was supported in part  by the DFG and NSFC through funds provided to the Sino-German
CRC 110 ``Symmetries and the Emergence of Structure in QCD'' (NSFC Grant No.  11621131001, DFG
Grant  No.    TRR110),  by  the  VolkswagenStiftung (Grant No.   93562), by 
the CAS President's International Fellowship Initiative (PIFI) (Grant No.   2018DM0034) and by the Georgian Shota Rustaveli National
Science Foundation (Grant No. FR17-354).


\end{document}